\documentclass[graybox]{svmult}

\usepackage{mathptmx}       % selects Times Roman as basic font
\usepackage{helvet}         % selects Helvetica as sans-serif font
\usepackage{courier}        % selects Courier as typewriter font
\usepackage{type1cm}        % activate if the above 3 fonts are
\usepackage{makeidx}         % allows index generation
\usepackage{graphicx}        % standard LaTeX graphics tool
\usepackage{multicol}        % used for the two-column index
\usepackage[bottom]{footmisc}% places footnotes at page bottom

\usepackage{amssymb}

\newcommand{\BEQ}{\begin{equation}}     % Gleichungen Anfang ..
\newcommand{\BEA}{\begin{eqnarray}}
\newcommand{\BD}{\begin{displaymath}}
\newcommand{\EEQ}{\end{equation}}       % .. und Ende
\newcommand{\EEA}{\end{eqnarray}}
\newcommand{\ED}{\end{displaymath}}
\newcommand{\eps}{\varepsilon}          % epsilon
\newcommand{\II}{{\rm i}}               % gerades i fuer komplexe Einheit
\newcommand{\demi}{\frac{1}{2}}         % Bruch 1/2
\newcommand{\wit}[1]{\widetilde{#1}}    % weite Schlange
\newcommand{\wht}[1]{\widehat{#1}}      % weiter Hut
\newcommand{\weps}{\wht{\,\epsilon\,}}
\newcommand{\vekz}[2]
     {\mbox{${\begin{array}{c} #1  \\ #2 \end{array}}$}}
\newcommand{\matz}[4] 
     {\mbox{${\begin{array}{cc} #1 & #2 \\ #3 & #4 \end{array}}$}}

\begin{document}

\title*{Physical ageing and new representations of some Lie algebras of local scale-invariance}
% Use \titlerunning{Short Title} for an abbreviated version of
% your contribution title if the original one is too long
\author{Malte Henkel and Stoimen Stoimenov}
% Use \authorrunning{Short Title} for an abbreviated version of
% your contribution title if the original one is too long
\institute{Malte Henkel \at Groupe de physique Statistique, 
Institut Jean Lamour (CNRS UMR 7198), Universit\'e de Lorraine Nancy, 
B.P. 70239, F - 54506 Vand{\oe}uvre-l\`es-Nancy Cedex, France, %\email{malte.henkel@univ-lorraine.fr}
\and Stoimen Stoimenov \at Institute of Nuclear Research and Nuclear Energy, 
Bulgarian Academy of Sciences, 
72 Tsarigradsko chaussee, Blvd., BG -- 1784 Sofia, Bulgaria %\email{name@email.address}
}
\maketitle

\abstract{Indecomposable but reducible representations of several Lie algebras of local scale-transformations,
including the Schr\"odinger and conformal Galilean algebras, 
and their applications in physical ageing are reviewed. The physical requirement of the decay of co-variant
two-point functions for large distances is related to analyticity properties in the coordinates dual to the
physical masses or rapidities.}

\section{Introduction}\label{sec:1}

Scale-invariance is recognised as one of the main characteristics of phase transitions and
critical phenomena. In addition, it has also become common folklore that given sufficiently local 
interactions, scale-invariance can be extended to larger Lie groups of coordinate transformations. 
Quite a few counter-examples exist, but the folklore carries on. Here, we are interested in the phenomenology of
phase transitions, either at equilibrium or far from equilibrium and shall study situations when scale-invariance
does indeed extend to conformal invariance or one of the generalisations appropriate for scale-invariant dynamics. 
We review recent results on indecomposable, but reducible (`logarithmic') representations and
discuss sufficient conditions which guarantee they decay of co-variant two-point functions at large distances. 

Consider the transformations in 
$(1+d)$-dimensional time-space $\mathbb{R}\otimes\mathbb{R}^d$
\BEQ
t \mapsto t'= \frac{\alpha t+\beta}{\gamma t +\delta} \;\; , \;\;
\vec{r}\mapsto \vec{r}' 
= \frac{{\cal R}\vec{r}+ \vec{v}_{2\ell} t^{2\ell} + \ldots \vec{v}_1 t + \vec{v}_0}{(\gamma t +\delta)^{2\ell}}
\;\; ; \;\; \alpha\delta-\beta\gamma=1
\EEQ
where ${\cal R}\in\mbox{\sl SO}(d)$, $\vec{v}_0,\ldots,\vec{v}_{2\ell}\in\mathbb{R}^d$ 
and $\alpha,\beta,\gamma,\delta\in\mathbb{R}$. 
The infinitesimal generators from these transformations 
only close in a Lie algebra if $\ell\in \demi\mathbb{N}$ (sometimes called `{\em spin-$\ell$ algebra}'). 
In $(1+1)$ dimensions, this algebra can be formulated in terms of two infinite families of generators 
$\mathfrak{sc}(1,\ell) := \langle X_n, Y_m\rangle_{n\in\mathbb{Z},m+\ell\in\mathbb{Z}}$
of the form
\BEQ \label{1.2}
X_n = -t^{n+1}\partial_t - (n+1) \ell t^n r \partial_r \;\; , \;\; 
Y_m = - t^{m+\ell} \partial_r 
\EEQ
with the non-vanishing commutators \cite{Henkel97}
\BEQ \label{1.3}
\left[ X_n, X_{n'} \right] = \left(n-n'\right) X_{n+n'} \;\; , \;\; 
\left[ X_n, Y_m\right] =\left( \ell n-m \right) Y_{n+m}
\EEQ
and where $z :=1/\ell$ is the\index{dynamical exponent} {\em dynamical exponent}. 
The maximal finite-dimensional sub-algebra is 
$\mathfrak{spi}(1,\ell):=\langle X_{\pm 1,0},Y_{-\ell,\ldots,\ell-1,\ell}\rangle$, 
for $\ell\in\demi\mathbb{N}$. 
In analogy with conformal invariance, Ward identities must be formulated
which will describe the action of these generators on scaling operators 
such that the co-variance under these transformations
can be used to derive differential equations to be satisfied by $n$-point correlators. 
Alternatively, one may include the corresponding terms directly 
into the generators themselves, which has the advantage that the verification
of the commutators guarantees the self-consistency of co-variance. 
However, the explicit representation (\ref{1.2}) does not take into account any
transformation properties of the (quasi-)primary scaling operators on which it is assumed to act. 

Dynamic time-space symmetries with a generic dynamical exponent ($z\ne 1$ possible) 
often arise as `non-relativistic limits' of the conformal
algebra. The two best-known examples are (i) the {\em Schr\"odinger algebra}\index{Schr\"odinger algebra} 
$\mathfrak{sch}(d)$ which in (\ref{1.2}) 
corresponds to $\ell=\demi$ (discovered in 1842/43 by Jacobi and in 1881 by Lie)
and (ii) the {\em conformal Galilean algebra}\index{conformal Galilean algebra} $\mbox{\sc cga}(d)$ \cite{Havas78} 
which corresponds in (\ref{1.2}) to $\ell=1$.  These two important
special cases can also be obtained by two distinct, complementary approaches
\begin{enumerate}
\item The non-relativistic limit of time-space conformal transformations such that a fixed value of the
dynamical exponent $z$ is assumed, reproduces the Schr\"odinger and conformal Galilean algebras from the
restriction to flat time-like and light-like geodesics, along with $z=2$ and $z=1$ \cite{Duval09}. 
\item If one tries to include the transformations of scaling operators into the representation 
(\ref{1.3}) of the infinite-dimensional 
algebra (\ref{1.3}) only the cases with $\ell=\demi, 1$ close as a Lie algebra. 
Besides the conformal algebra, this
reproduces the Schr\"odinger and conformal Galilean algebras \cite{Henkel02}. 
\end{enumerate}

Including the terms which describe the transformation of quasi-primary scaling operators 
often leads to central extensions of the above algebras. For $\ell=\demi$, 
one has instead of $\mathfrak{spi}(1,\demi)$ the
{\em Schr\"odinger-Virasoro algebra}\index{Schr\"odinger-Virasoro algebra} 
$\mathfrak{sv}=\langle X_n, Y_m, M_n\rangle_{n\in\mathbb{Z},m\in\mathbb{Z}+\demi}$ \cite{Henkel94,Unterberger12}
spanned by the generators
\BEA
X_n  &=& - t^{n+1}\partial_t - \frac{n+1}{2} t^n r\partial_r 
- \frac{n+1}{2} x t^n - \frac{n(n+1)}{4}{\cal M} t^{n-1} r^2 
\nonumber \\
Y_m &=& - t^{m+\demi} \partial_r - \left(m+\demi\right){\cal M} t^{m-\demi} r
\;\;\;\; , \;\;\;\;  M_n = - t^n {\cal M} \label{1.4} 
\EEA
($x$ is the scaling dimension and $\cal M$ the mass)\index{mass} and the non-vanishing commutators
\BEA
\left[ X_n, X_{n'}\right] = (n-n')X_{n+n'} &\;\;,\;\;& 
\left[ X_n, Y_m\right] = \left(\frac{n}{2}-m\right) Y_{n+m}
\nonumber \\ 
\left[ X_n, M_{n'} \right] = -n' M_{n+n'} &\;\; , \;\;& 
\left[Y_m, Y_{m'} \right] = \left(m-m'\right) M_{m+m'}
\EEA 
Its maximal finite-dimensional sub-algebra is the Schr\"odinger algebra 
$\mathfrak{sch}(1)=\langle X_{\pm 1,0}, Y_{\pm\demi}, M_0\rangle$, 
which centrally extends $\mathfrak{spi}(1,\demi)$. 
It is the maximal dynamical symmetry of the free Schr\"odinger equation
${\cal S}\phi=0$ with\index{Schr\"odinger operator}
 ${\cal S}=2M_0 X_{-1} - Y_{-\demi}^2=2{\cal M}\partial_t - \partial_r^2$, 
in the sense that solutions of ${\cal S}\phi=0$ with scaling dimension 
$x=x_{\phi}=\demi$ are mapped onto solutions, 
a fact already known to Jacobi and to Lie. 
More generally, unitarity implies the bound $x\geq \demi$ \cite{Lee09}. 

On the other hand, for $\ell=1$ one obtains the {\em altern-Virasoro algebra} 
$\mathfrak{av}:=\langle X_n, Y_n\rangle_{n\in\mathbb{Z}}$ (also called `full \mbox{\sc cga}'), 
with an explicit representation spanned by 
\cite{Havas78,Henkel97,Negro97,Henkel02,Galajinsky08,Bagchi09,Bagchi10,Martelli09,Cherniha10,Hosseiny10b,Hotta10,Aizawa12,Chakraborty12,Lukierski12,Andr13,Kim13,Henkel14}
\BEA \label{1.6}
X_n &=& -t^{n+1} \partial_t - (n+1)t^n r\partial_r - (n+1)x t^n -n(n+1)\gamma t^{n-1} r \nonumber \\
Y_n &=& - t^{n+1} \partial_r - (n+1) \gamma t^n
\EEA
which obeys (\ref{1.3}) and has $\mbox{\sc cga}(1)=\langle X_{\pm 1,0},Y_{\pm 1,0}\rangle$ 
as maximal finite-dimensional sub-algebra.\footnote{In the context of asymptotically flat $3D$ gravity, 
an isomorphic Lie algebra is known as BMS algebra, $\mathfrak{bms}_3\equiv \mbox{\sc cga}(1)$ 
\cite{Barnich07,Barnich13,Bagchi12,Bagchi13b}.} 
The representation (\ref{1.6}) is spanned by the two scalars $x$ and $\gamma$. 

%%+++++++++++++++++++++++++++++++++++++++++++++++++++++++++++++++++++++++++++++++++++++++++++++++++++++++++++++++
\begin{figure}[t]
%\sidecaption[t]
\includegraphics[scale=.6]{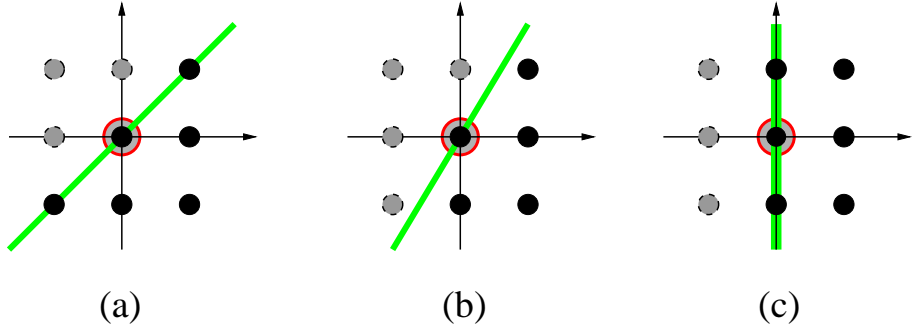}
\caption[fig1]{Root diagrammes of (a) $\mathfrak{sch}(1)$ (b) $\mathfrak{age}(1)$ and 
(c) $\mathfrak{alt}(1)=\mbox{\sc cga}(1)$ as sub-algebras
of the complex Lie algebra $B_2$. If the second generator in the centre is included (double red circle)
one obtains the maximal parabolic sub-algebras of $B_2$. 
\label{fig1} }
\end{figure}
%%+++++++++++++++++++++++++++++++++++++++++++++++++++++++++++++++++++++++++++++++++++++++++++++++++++++++++++++++

The relationship between $\mathfrak{sch}(1)$ and $\mbox{\sc cga}(1)$ can in be understood in a different way by
considering the imbedding $\mathfrak{sch}(1)\subset B_2$ into the complex Lie algebra $B_2$. This can be
visualised in terms of a root diagramme, see figure~\ref{fig1}(a), where the generators of $\mathfrak{sch}(1)$
are indicated by full black circles and the remaining ones by the grey circles. As it is well-known \cite{Knapp86},
a {\em standard parabolic sub-algebra}\index{parabolic sub-algebra} 
$\mathfrak{p}$, of a semi-simple Lie algebra $\mathfrak{g}$ consists 
of the Cartan sub-algebra $\mathfrak{h}\subset\mathfrak{g}$ and of all `positive' generators in $\mathfrak{g}$. 
The meaning of `positive' can be simply illustrated in figure~\ref{fig1} for the special case 
$\mathfrak{g}=B_2$: one draws a straight line through 
the centre of the root diagramme and all generators on that line or to the
right of it are `positive'. From figure~\ref{fig1}(a), one also sees that the Schr\"odinger algebra
can be extended to a parabolic sub-algebra 
$\wit{\mathfrak{sch}}(1)=\mathfrak{sch}(1)\oplus\mathbb{C}N$ by adding
an extra generator $N$, which is indicated by the red double circle in the centre. 
Since the Weyl reflections and rotations can be used to map isomorphic sub-algebras onto each other, 
the classification of the maximal parabolic sub-algebras of $B_2$ can now be illustrated simply through
the value of the slope $p$ of the straight line in figure~\ref{fig1} \cite{Henkel03}: 
\begin{enumerate}
\item if $p=1$, one has $\wit{\mathfrak{sch}}(1)=\mathfrak{sch}(1)+\mathbb{C}N$, 
the parabolic extension of the Schr\"odinger algebra, see
figure~\ref{fig1}(a). See below for explicit forms for $N$.  
\item if $1<p<\infty$, one has $\wit{\mathfrak{age}}(1)=\mathfrak{age}(1)+\mathbb{C}N$, 
the parabolic extension of the {\em ageing algebra}, 
see figure~\ref{fig1}(b).
\item if $p=\infty$, one has $\wit{\mbox{\sc cga}}(1)=\mbox{\sc cga}(1)+\mathbb{C}N$, 
the parabolic extension of the conformal Galilean algebra, see figure~\ref{fig1}(c).
\end{enumerate}  

While \mbox{\sc cga}(1) does not have a central extension, this is different in $d=2$ space dimensions, where
a so-called `exotic' central extension exists. 
This gives the {\em exotic conformal Galilean algebra}\index{exotic conformal Galilean algebra}
$\mbox{\sc ecga}=\langle X_{\pm 1,0}, Y_{\pm 1,0}, \theta, R_0\rangle$ 
\cite{Lukierski06,Lukierski07} with an explicit representation
(where $j,k,\ell=1,2$ and summation over repeated indices is implied) 
\BEA
X_n &=& - t^{n+1}\partial_t - (n+1) t^n {r_j}{\partial_j} - x (n+1) t^n -n (n+1) t^{n-1} \gamma_j r_j
- n(n+1) {h_j r_j} 
\nonumber \\
Y_n^{(j)} &=& - t^{n+1}\partial_j -(n+1) t^n \gamma_j - (n+1) t^n h_j -n(n+1)\theta \eps_{jk} r_k
\label{1.7} \\
J &=& - \eps_{k\ell}\, r_k \partial_{\ell} - 
\eps_{k\ell}\,\gamma_k \frac{\partial}{\partial \gamma_{\ell}} - \frac{1}{2\theta} {h_j}{h_j}
\nonumber
\EEA
characterised by a scalar scaling dimension $x$ and a vector $\vec{\gamma}=(\gamma_1,\gamma_2)$ 
of {\em `rapidities'}\index{rapidity} \cite{Martelli09,Cherniha10,Henkel14}. The
components of the vector $\vec{h}=(h_1,h_2)$ satisfy $[h_i,h_j]=\eps_{ij}\theta$, 
where $\theta$ is central. $\eps$ is the totally antisymmetric $2\times 2$ tensor and $\eps_{12}=1$. 
The non-vanishing commutators of the \mbox{\sc ecga} read 
\BEA
\left[ X_n, X_m \right] = (n-m) X_{n+m} &\;\;,\;\;&
\left[ X_n, Y_m^{(i)}\right] = (n-m) Y_{n+m}^{(i)} \nonumber \\
\left[ Y_n^{(i)}, Y_m^{(j)} \right] = \eps_{ij} \, \delta_{n+m,0} \left( 3\delta_{n,0} - 2\right) \theta 
&\;\;,\;\;& 
\left[ J, Y_n^{(i)}\right] = \eps_{i j} Y_n^{(j)} 
\label{1.8}
\EEA 
and the \mbox{\sc ecga}-invariant Schr\"odinger operator\index{Schr\"odinger operator} is
\BEQ \label{1.9}
{\cal S} = -\theta X_{-1} + \eps_{ij} Y_0^{(i)} Y_{-1}^{(j)} 
= \theta \partial_t + \eps_{ij} \left( \gamma_i + h_i \right) \partial_j
\EEQ
with $x=x_{\phi}=1$. The unitary bound gives $x\geq 1$ \cite{Martelli09}. 

The common sub-algebra of $\mathfrak{sch}(1)$ and $\mbox{\sc cga}(1)$ 
is called the {\em ageing algebra}\index{ageing algebra}
$\mathfrak{age}(1):=\langle X_{0,1},Y_{\pm\demi},M_0\rangle$ and does not include time-translations. 
Starting from the representation (\ref{1.4}), only the
generators $X_n$ assume a more general form \cite{Henkel06}  
\BEQ \label{1.10}
X_n  = - t^{n+1}\partial_t - \frac{n+1}{2} t^n r\partial_r - \frac{n+1}{2} x t^n 
- n(n+1)\xi t^{n} -\frac{n(n+1)}{4}{\cal M} t^{n-1} r^2 
\EEQ
which also admits a more general invariant Schr\"odinger operator\index{Schr\"odinger operator}
${\cal S}=2{\cal M}\partial_t - \partial_r^2 +2{\cal M}t^{-1}\left(x+\xi-\demi\right)$, 
but without any constraint on  neither $x$ nor $\xi$ \cite{Stoimenov13}. 
This representation of $\mathfrak{age}(1)$ is characterised by the scalars 
$(x,\xi,{\cal M})$. The name of this algebra comes from
its use as dynamical symmetry in {\em physical ageing},\index{ageing}\index{\physical ageing} 
which can be observed in strongly interacting many-body systems quenched
from a disordered initial state to the co-existence 
regime below the critical temperature $T_c>0$ where several equivalent equilibrium
states exist. For example, for quenched Ising spins in $d\geq 2$ dimensions 
without disorder, nor frustrations, and with a purely relaxational dynamics without any conservation law, 
it can be shown that the dynamical exponent $z=2$ \cite{Bray94}. 
Assuming $\mathfrak{age}(d)$ as a dynamical symmetry predicts 
the form of the two-time linear response function of the average
order parameter $\langle \phi(t,\vec{r})\rangle$ 
with respect to its canonically conjugate magnetic field 
$h(s,\vec{r}')$\index{response function}\index{two-point function}
\cite{Henkel94,Henkel03,Henkel06}
\BEA
R(t,s;\vec{r}) &=& \left.\frac{\delta \langle \phi(t,\vec{r})\rangle}{\delta h(s,\vec{0})}\right|_{h=0} = 
\left\langle \phi(t,\vec{r}) \wit{\phi}(s,\vec{0}) \right\rangle = 
s^{-1-a} F_R\left(\frac{t}{s},\frac{\vec{r}^2}{t-s}\right)
\label{1.11} \\
F_R(y,u) &=& F_0\, \delta({\cal M}-\wit{\cal M\!}\:)\, 
\Theta(y-1)\, y^{1+a'-\lambda_R/z} (y-1)^{-1-a'} \exp\left[{-\demi {\cal M} u}\right] 
\nonumber
\EEA
where the standard Janssen-de Dominicis formalism (see e.g. \cite{Henkel10}) was used to re-write 
$R=\langle \phi\wit{\phi}\rangle$ as a correlator of the order parameter
scaling operator $\phi$ and the conjugate response operator $\wit{\phi}$.\index{response operator} 
Both of these are assumed to be quasi-primary under $\mathfrak{age}(d)$. 
The ageing exponents $a,a',\lambda_R/z$ are related to
$x,\xi$ and $\wit{x},\wit{\xi}$ in a known way, e.g. $a'-a=\frac{2}{z}(\xi+\wit{\xi})$. 
$F_0$ is a normalisation constant and the $\Theta$-function expresses the causality condition
$y=t/s>1$, of which we shall say more in section~\ref{sec:3} below. Spatial translation-invariance was assumed. 

The case of a Schr\"odinger-invariance response is obtained if one sets 
$\xi=\wit{\xi}=0$, hence $a=a'$ in (\ref{1.11}). 

Eq.~(\ref{1.11}) has been confirmed in numerous spin systems 
(e.g. Ising, Potts, XY, spherical, Hilhorst-van Leeuven, Edwards-Wilkinson,\ldots models) 
which undergo simple ageing with $z=2$, 
both for the time- and space-dependence; 
either from a known exact solution or using simulational data. 
For a detailed review, see \cite{Henkel10}. 
Current empirical evidence suggests that for quenches to low temperatures $T<T_c$, 
one should have for the second scaling dimensions $\xi+\wit{\xi}=0$, hence $a=a'$. 
However, the full representation (\ref{1.10}) of $\mathfrak{age}(1)$ 
is needed in the $d=1$ Glauber-Ising model, where
the exact solution reproduces (\ref{1.11}) with $a=0$, $a'-a=-\demi$ 
and $\lambda_R=1$ \cite{Henkel06}.\index{ageing algebra} 
One might anticipate that $a'-a\ne 0$ for quenches to the critical point $T=T_c$.  

For critical quenches, one has in general $z\ne 2$, such that (\ref{1.11}) 
does no longer apply. However, the form of the auto-response
$R(t,s)=R(t,s;\vec{0})$ does not contain the precise 
spatial form so that at least that part of (\ref{1.11}) can be used for preliminary tests
of dynamical symmetries for generic values of $z$. 

In section~\ref{sec:2}, various logarithmic representations of these algebras and some of their
properties are reviewed. Known applications to physical ageing will be briefly discussed. In section~\ref{sec:3}, 
the requirement of a physically sensible limit in the case of large spatial separation 
$|\vec{r}_1-\vec{r}_2|\to\infty$ leads to the derivation of causality conditions. These inform
on the interpretation in terms of either responses or correlators.

\section{Logarithmic representations}\label{sec:2}

Logarithmic\index{logarithmic conformal invariance} conformal field-theories arise from
indecomposable but reducible representations of the Virasoro algebra
\cite{Saleur92,Gurarie93,Mathieu07,Ruelle13}, see \cite{Ridout14} 
for a collection of recent reviews. Formally, in the most simple case, 
this can be implemented \cite{Gurarie93,Rahimi97} by replacing the order parameter
$\phi$ by a vector {\small $\Phi=\left(\vekz{\psi}{\phi}\right)$} such that the scaling
dimension $x$ in the Lie algebra generators becomes a Jordan matrix {\small $\left(\matz{x}{1}{0}{x}\right)$}. 
Anti\-ci\-pa\-ting the notation used for the algebras we are going to consider, we introduce, instead of a single
two-point function $\langle \phi_1\phi_2\rangle$, the three two-point functions\index{two-point function} 
\BEQ \label{2.12}
F := \left\langle \phi_1(t_1,\vec{r}_1) \phi_2^*(t_2,\vec{r}_2)\right\rangle \; , \; 
G := \left\langle \phi_1(t_1,\vec{r}_1) \psi_2^*(t_2,\vec{r}_2)\right\rangle \; , \; 
H := \left\langle \psi_1(t_1,\vec{r}_1) \psi_2^*(t_2,\vec{r}_2)\right\rangle
\EEQ 
Temporal and spatial translation-invariance imply that
$F=F(t,\vec{r})$, $G=G(t,\vec{r})$ and $H=H(t,\vec{r})$ with $t=t_1-t_2$ and
$\vec{r}=\vec{r}_1-\vec{r}_2$. The shape of these functions is derived from the algebras
introduced in section~\ref{sec:1}, as we now review.

\subsection{Schr\"odinger algebra}

For the Schr\"odinger and the ageing algebras, the `complex conjugate' $\phi^*$ in (\ref{2.12}) 
refers to the mapping ${\cal M}\mapsto -\wit{\cal M\!}$\index{Bargman superselection rule}\index{mass} 
when in a response function such as (\ref{1.11}) one goes from the order parameter 
$\phi$ to its conjugate response operator\index{response operator} 
$\wit{\phi}$ \cite{Henkel10}. This is necessary in applications to physical ageing. 
In particular, extending
$\mathfrak{sch}(1)\to\wit{\mathfrak{sch}}(1)$ and using the physical convention 
${\cal M}\geq 0$ of non-negative masses, 
implies causality $t_1-t_2>0$, as we shall see in section~\ref{sec:3}.
While common in statistical physics applications in models described by stochastic Langevin
equations \cite{Henkel06}, this was recently re-discovered in string-theory contexts \cite{Nakayama10}. 

Replacing in the generators (\ref{1.4}) the scaling dimension $x$ by a $2\times 2$\index{Jordan matrix} 
Jordan matrix, the Schr\"odinger Ward identities (or co-variance conditions)
can be written down for the three two-point functions (\ref{2.12}).
The result is, in $d\geq 1$ dimensions \cite{Hosseiny10}\index{Schr\"odinger algebra}
%\newpage \typeout{*** saut de page ***}
\BEA
F(t,\vec{r}) = \langle\phi_1(t,\vec{r})\phi_2^*(0,\vec{0})\rangle &=& 0, \nonumber \\
G(t,\vec{r}) = \langle\phi(t,\vec{r})\psi_2^*(0,\vec{0})\rangle &=&
a\, t^{-2x_1}\,\exp\left[{-\frac{{\cal{M}}_1\vec{r}^2}{2t}}\right]\,
\delta_{x_1,x_2}\delta_{{\cal{M}}_1,{\cal{M}}_2}, 
\label{2.13} \\ 
H(t,\vec{r}) = \langle\psi_1(t,\vec{r})\psi_2^*(0,\vec{0})\rangle &=& 
t^{-2x_1}\,\left(b-2a\ln{t}\right)\,\exp\left[{-\frac{{\cal{M}}_1\vec{r}^2}{2t}}\right]\, 
\delta_{x_1,x_2}\delta_{{\cal{M}}_1,{\cal{M}}_2}.
\nonumber
\EEA
where $a,b$ are scalar normalisation constants. Time- and space-translation-invariance and 
also rotation-invariance for scalar $\phi,\psi$ were used. 

\subsection{Conformal Galilean algebra}

The representation (\ref{1.6}) of $\mbox{\sc cga}(1)$ 
depends on both the scaling dimension $x$ \cite{Hosseiny11} as well as the rapidity $\gamma$. 
Now, either of them may become a Jordan matrix and it turns out that the 
$\mbox{\sc cga}(1)$-commutators imply that 
{\it simultaneously} \cite{Henkel14}\index{conformal Galilean algebra}\index{Jordan matrix}
\BEQ \label{2.14}
x \mapsto \left(\matz{x}{x'}{0}{x}\right) \;\; , \;\; 
\gamma\mapsto \left(\matz{\gamma}{\gamma'}{0}{\gamma}\right)
\EEQ
In  contrast to $\mathfrak{sch}(d)$, the `complex conjugate' 
is not needed here. The \mbox{\sc cga}-Ward identities lead to
\cite{Henkel14}, immediately written down for $d\geq 1$ spatial dimensions
\BEA
F = \langle \phi_1 \phi_2\rangle(t,\vec{r}) &=& 0 \nonumber \\
G = \langle \phi_1 \psi_2\rangle(t,\vec{r}) &=& a |t|^{-2x_1} e^{-2\vec{\gamma}_1\cdot\vec{r}/t} \: 
\delta_{x_1,x_2}\delta_{\vec{\gamma}_1,\vec{\gamma}_2}\, 
\delta_{x'_1,x'_2}\delta_{\vec{\gamma'}_1,\vec{\gamma'}_2}
\label{2.15} \\
H = \langle \psi_1 \psi_2\rangle(t,\vec{r}) &=& |t|^{-2x_1} e^{-2\vec{\gamma}_1\cdot\vec{r}/t} 
\left[ b -2a \frac{\vec{r}}{t}\cdot\vec{\gamma'}_1 -2a x_1' \ln|t|  \right]
\delta_{x_1,x_2}\delta_{\vec{\gamma}_1,\vec{\gamma}_2} 
\delta_{x'_1,x'_2}\delta_{\vec{\gamma'}_1,\vec{\gamma'}_2}
\nonumber
\EEA
where the normalisation $a=a(\vec{\gamma}_1^2,{\vec{\gamma}_1'}^2,\vec{\gamma}_1\cdot\vec{\gamma}_1')$ 
as follows from rotation-invariance for $d>1$ and an analogous form holds for $b$.

\subsection{Exotic conformal Galilean algebra}

Again, both the scalar $x$ as well as the vector $\vec{\gamma}$ 
may become simultaneously Jordan matrices, according to (\ref{2.14}). 
One then needs four distinct two-point functions\index{exotic conformal Galilean algebra} 
\BEQ \label{2.16}
F = \langle \phi_1 \phi_2\rangle \;\; , \;\;
G_{12} = \langle \phi_1 \psi_2\rangle \;\; , \;\;
G_{21} = \langle \psi_1 \phi_2\rangle \;\; , \;\;
H = \langle \psi_1 \psi_2\rangle
\EEQ
which all depend merely on $t=t_1-t_2$ and $\vec{r}=\vec{r_1}-\vec{r}_2$. 
The operators $\vec{h},\theta$ 
are realised in terms of auxiliary
variables $\vec{\nu}$ such that $h_i = \partial_{\nu_i}-\demi \epsilon_{ij}\nu_j\theta$ 
with $i,j=1,2$. 
Remarkably, it turns out that that two cases must be distinguished \cite{Henkel14}:\\

\noindent \underline{\bf Case 1}: {\it defined by $x_1'\ne0$ or $x_2'\ne 0$ and $F=0$.} \\
In what follows, the indices always refer to the identity 
of the two primary operators $\Phi_{1,2}=\left(\vekz{\psi_{1,2}}{\phi_{1,2}}\right)$.
We also use the two-dimensional vector product (with a scalar value) 
$\vec{a}\wedge\vec{b} := \epsilon_{ij} a_i b_j$.  
Then  $G_{12}=G(t,\vec{r})=G(-t,-\vec{r})=G_{21}=:G$ 
such that one has the constraints $x_1=x_2$, $x_1'=x_2'$,  $\theta_1+\theta_2=0$ and (recall $d=2$)
\BEA
\hspace{-0.3truecm}G &=& |t|^{-2x_1} 
e^{-(\vec{\gamma}_1+\vec{\gamma}_2)\cdot\vec{u} 
-\demi (\vec{\gamma}_1-\vec{\gamma}_2)\cdot(\vec{\nu}_1-\vec{\nu}_2)}\, 
e^{\theta_1 \vec{u}\wedge(\vec{\nu}_1-\vec{\nu}_2) +\demi\theta_1 \vec{\nu}_1\wedge\vec{\nu}_2}\,
g_0(\vec{w}) 
\nonumber \\
\hspace{-0.3truecm}H &=& |t|^{-2x_1} 
e^{-(\vec{\gamma}_1+\vec{\gamma}_2)\cdot\vec{u} 
-\demi (\vec{\gamma}_1-\vec{\gamma}_2)\cdot(\vec{\nu}_1-\vec{\nu}_2)}\, 
e^{\theta_1 \vec{u}\wedge(\vec{\nu}_1-\vec{\nu}_2) +\demi\theta_1 \vec{\nu}_1\wedge\vec{\nu}_2}\,
h(\vec{u},\vec{\nu}_1,\vec{\nu}_2) 
\label{ecga_case1} \\
\hspace{-0.3truecm}h &=& h_0(\vec{w}) 
- g_0(\vec{w}) \left( 2x_1' \ln|t| + \vec{u}\cdot\left(\vec{\gamma}_1'+\vec{\gamma}_2'\right)
+\demi\left(\vec{\nu}_1-\vec{\nu}_2\right)\cdot\left(\vec{\gamma}_1'-\vec{\gamma}_2'\right) \right)
\nonumber
\EEA
together with the abbreviations $\vec{u}=\vec{r}/t$ and 
$\vec{w} := \vec{u}-\demi\left(\vec{\nu}_1+\vec{\nu}_2\right)$. The functions $g_0(\vec{w})$ 
and $h_0(\vec{w})$ remain undetermined.  

\noindent \underline{\bf Case 2}: {\it defined by $x_1'=x_2'= 0$}, 
hence only the vector $\vec{\gamma}$ has a Jordan form. \\
One has the constraints $x_1=x_2$, $\theta_1+\theta_2=0$ and
\BEA
\hspace{-0.6truecm}F &=& |t|^{-2x_1} 
e^{-(\vec{\gamma}_1+\vec{\gamma}_2)\cdot\vec{u} 
-\demi (\vec{\gamma}_1-\vec{\gamma}_2)\cdot(\vec{\nu}_1-\vec{\nu}_2)}\, 
e^{\theta_1 \vec{u}\wedge(\vec{\nu}_1-\vec{\nu}_2) +\demi\theta_1 \vec{\nu}_1\wedge\vec{\nu}_2}\,
f_0(\vec{w}) 
\nonumber \\
\hspace{-0.6truecm}G_{12} &=& |t|^{-2x_1} 
e^{-(\vec{\gamma}_1+\vec{\gamma}_2)\cdot\vec{u} 
-\demi (\vec{\gamma}_1-\vec{\gamma}_2)\cdot(\vec{\nu}_1-\vec{\nu}_2)}\, 
e^{\theta_1 \vec{u}\wedge(\vec{\nu}_1-\vec{\nu}_2) +\demi\theta_1 \vec{\nu}_1\wedge\vec{\nu}_2}\,
g_{12}(\vec{u},\vec{\nu}_1,\vec{\nu}_2) 
\nonumber \\
\hspace{-0.6truecm}G_{21} &=& |t|^{-2x_1} 
e^{-(\vec{\gamma}_1+\vec{\gamma}_2)\cdot\vec{u} 
-\demi (\vec{\gamma}_1-\vec{\gamma}_2)\cdot(\vec{\nu}_1-\vec{\nu}_2)}\, 
e^{\theta_1 \vec{u}\wedge(\vec{\nu}_1-\vec{\nu}_2) +\demi\theta_1 \vec{\nu}_1\wedge\vec{\nu}_2}\,
g_{21}(\vec{u},\vec{\nu}_1,\vec{\nu}_2) 
\label{ecga_case2}\\
\hspace{-0.6truecm}H &=& |t|^{-2x_1} 
e^{-(\vec{\gamma}_1+\vec{\gamma}_2)\cdot\vec{u} 
-\demi (\vec{\gamma}_1-\vec{\gamma}_2)\cdot(\vec{\nu}_1-\vec{\nu}_2)}\, 
e^{\xi_1 \vec{u}\wedge(\vec{\nu}_1-\vec{\nu}_2) +\demi\theta_1 \vec{\nu}_1\wedge\vec{\nu}_2}\,
h(\vec{u},\vec{\nu_1},\vec{\nu}_2) 
\nonumber 
\EEA
where 
\BEA
g_{12} &=&  g_{0}(\vec{w}) - f_0(\vec{w}) 
\left( \vec{u}-\demi\left(\vec{\nu}_1-\vec{\nu}_2\right)\right)\cdot\vec{\gamma}_2' \nonumber \\
g_{21} &=&  g_{0}(\vec{w}) - f_0(\vec{w}) 
\left( \vec{u}+\demi\left(\vec{\nu}_1-\vec{\nu}_2\right)\right)\cdot\vec{\gamma}_1'  
\label{ecga_gh}\\
h      &=& h_0(\vec{w}) - g_0(\vec{w}) \left( \vec{u}\cdot\left(\vec{\gamma}_1'+\vec{\gamma}_2'\right)
+\demi\left(\vec{\nu}_1-\vec{\nu}_2\right)\cdot\left(\vec{\gamma}_1'-\vec{\gamma}_2'\right) \right)
\nonumber \\
& & +\demi f_0(\vec{w}) \left( \vec{u}+\demi\left(\vec{\nu}_1-\vec{\nu}_2\right)\right)\cdot\vec{\gamma}_1' \: 
\left( \vec{u}-\demi\left(\vec{\nu}_1-\vec{\nu}_2\right)\right)\cdot\vec{\gamma}_2'
\nonumber
\EEA
The functions $f_0(\vec{w})$, $g_0(\vec{w})$ 
and $h_0(\vec{w})$ remain undetermined. 

Finally, two distinct choices for the rotation generator have been considered in the
litt\'erature, namely (single-particle form)\index{rotation-invariance} 
\BEQ
J = -\vec{r}\wedge\partial_{\vec{r}} - \vec{\gamma} \wedge \partial_{\vec{\gamma}} 
- \frac{1}{2\theta} \vec{h}\cdot\vec{h}
\;\; \mbox{\rm ~and~} \;\;
R = -\vec{r}\wedge\partial_{\vec{r}} - \vec{\gamma} \wedge \partial_{\vec{\gamma}} 
- \vec{\nu}\wedge \partial_{\vec{\nu}}
\EEQ
The generator $J$ arises naturally when one derives the generators of the 
\mbox{\sc ecga} from a contraction of a pair conformal
algebras with non-vanishing spin \cite{Martelli09}, whereas the choice 
$R$ has a fairly natural form, especially in the auxiliary variables
$\vec{\nu}$. Both generators obey the same commutators (\ref{1.8}) 
with the other generators of {\sc ecga} and commute with the 
Schr\"odinger operator (\ref{1.9}). 
One speaks of {\em `$J$-invariance'} if the generator $J$ 
is used and of {\em `$R$-invariance'},  if the generator $R$ is used. 
The consequences of both cases are different \cite{Henkel14}: 

\noindent {\bf A)} If one uses $R$-invariance, 
in both cases the functions $f_0(\vec{w})$, $g_{0}(\vec{w})$ 
and $h_0(\vec{w})$ are short-hand notations for undetermined functions of 9 
rotation-invariant combinations of $\vec{w}$, $\vec{\gamma}_{1,2}$ and
$\vec{\gamma}_{1,2}'$, for example
\BEQ \label{2.21}
f_0=f_0\left( \vec{w}^2,\vec{\gamma}_1^2,\vec{\gamma}_2^2,{\vec{\gamma}_1'}^2, 
{\vec{\gamma}_2'}^2,\vec{w}\cdot\vec{\gamma}_1,\vec{w}\cdot\vec{\gamma}_2,
\vec{w}\cdot\vec{\gamma}_1',\vec{w}\cdot\vec{\gamma}_2'\right)
\EEQ
and analogously for $g_0$ and $h_0$.  \\
{\bf B)} For $J$-invariance, the $\vec{\gamma}$-matrices become diagonal, 
viz. $\vec{\gamma}_1'=\vec{\gamma}_2'=\vec{0}$. Then 
only case 1 retains a logarithmic (i.e. indecomposable) structure and, 
with {\small $\weps=\left(\matz{0}{1}{-1}{0}\right)$}  
\BEA 
g_0=g_0\left( \vec{\gamma}_1^2,\vec{\gamma}_2^2,\vec{\gamma}_1\cdot\vec{\gamma}_2, 
\vec{w}+\weps \left(\vec{\gamma}_1-\vec{\gamma}_2\right) (2\theta_1)^{-1} \right)
\nonumber \\
h_0=h_0\left( \vec{\gamma}_1^2,\vec{\gamma}_2^2,\vec{\gamma}_1\cdot\vec{\gamma}_2, 
\vec{w}+\weps \left(\vec{\gamma}_1-\vec{\gamma}_2\right) (2\theta_1)^{-1} \right)
\label{2.22}
\EEA

In all these {\sc ecga}-covariant two-point functions, there  never is a constraint on the
$\vec{\gamma}_i$, and on the $\vec{\gamma}_i'$ only in the case of $J$-invariance. 

\subsection{Ageing algebra}

Since the representation (\ref{1.10}) of $\mathfrak{age}(1)$ 
contains the two independent scaling dimensions $x,\xi$, either
of those may take a matrix form. From the commutators, 
it can be shown that both are simultaneously of Jordan form 
\cite{Henkel13}\index{ageing algebra}\index{Jordan matrix}
\BEQ \label{2.23}
x \mapsto \left(\matz{x}{x'}{0}{x}\right) \;\; , \;\; 
\xi\mapsto \left(\matz{\xi}{\xi'}{0}{\xi}\right)
\EEQ
We use again the definitions (\ref{2.16}). 
Since the space-dependent part of the two-point functions has the same form
as already derived above, in the case of Schr\"odinger-invariance, 
we set $\vec{r}=\vec{r}_1-\vec{r}_2=\vec{0}$ and consider only the
time-dependent part where only the values of the exponents change when 
$z\ne 2$ is admitted. The requirement of co-variance under
this logarithmic representation of $\mathfrak{age}(1)$ leads to \cite{Henkel13}
\BEA
F(t,s) &=& s^{-(x_1+x_2)/2}\: {\cal F}\left(\frac{t}{s}\right)  f_0 
\nonumber \\
G_{12}(t,s) &=& s^{-(x_1+x_2)/2}\: {\cal F}\left(\frac{t}{s}\right) 
\left( g_{12}\left(\frac{t}{s}\right) +  \gamma_{12}\left(\frac{t}{s}\right)\ln s \right) 
\nonumber \\
G_{21}(t,s) &=& s^{-(x_1+x_2)/2}\: {\cal F}\left(\frac{t}{s}\right)
\left( g_{21}\left(\frac{t}{s}\right) +  \gamma_{21}\left(\frac{t}{s}\right)\ln s  \right)   
\label{2.24}  \\
H(t,s) &=& s^{-(x_1+x_2)/2} \: {\cal F}\left(\frac{t}{s}\right)
\left( h_0\left(\frac{t}{s}\right) 
+ h_1\left(\frac{t}{s}\right)\ln s  + h_2\left(\frac{t}{s}\right)\ln^2 s \right) \nonumber
\EEA
with the abbreviation ${\cal F}(y)=y^{2\xi_2/z +(x_2-x_1)/z} (y-1)^{-(x_1+x_2)/z-2(\xi_1+\xi_2)/z}$. 
Herein the scaling functions, depending only on $y=t/s$, are given by 
\BEA 
\gamma_{12}(y)  = -\demi x_2' f_0 &\;\; , \;\;& \gamma_{21}(y) = -\demi x_1' f_0
\nonumber \\
h_1(y) = -\demi \left( x_1' g_{12}(y) + x_2' g_{21}(y) \right) &\;\; , \;\;& 
h_2(y) = \frac{1}{4} x_1' x_2' f_0 
\label{2.25}
\EEA
and
\BEA
g_{12}(y) &=& g_{12,0} +\left(\frac{x_2'}{2}+\xi_2'\right) f_0 \ln \left|\frac{y}{y-1}\right| 
\nonumber \\
g_{21}(y) &=& g_{21,0} -\left(\frac{x_1'}{2}+\xi_1'\right) f_0 \ln |y-1| - \frac{x_1'}{2} f_0 \ln |y| 
\nonumber \\
h_0(y) &=& h_0 - \left[ \left(\frac{x_1'}{2}+\xi_1'\right)g_{21,0} + 
\left(\frac{x_2'}{2}+\xi_2'\right)g_{12,0}\right]\ln|y-1| \nonumber \\
& & - \left[ \frac{x_1'}{2} g_{21,0} - \left(\frac{x_2'}{2}+\xi_2'\right)g_{12,0}\right]\ln|y| 
\label{2.26} \\
& &  + \demi f_0 \left[ \left( \left(\frac{x_1'}{2} +\xi_1'\right)\ln |y-1| 
+ \frac{x_1'}{2}\ln |y|\right)^2 
-  \left(\frac{x_2'}{2} +\xi_2'\right)^2 \ln^2\left|\frac{y}{y-1}\right| \right] 
\nonumber 
\EEA
and $f_0, g_{12,0}, g_{21,0}, h_0$ are normalisation constants.
There are no constraints on any of the $x_i, x_i', \xi_i,\xi_i'$. 
However, for $z=2$ the Bargman superselection rule
${\cal M}_1+{\cal M}_2=0$ holds true.  

\subsection{Discussion} 

Table~\ref{tab1} summarises some features of the 
various logarithmic representations considered here.\index{Bargman superselection rule} 
First, logarithmic Schr\"odinger-invariance, 
with a single scaling dimension $x$ elevated to a Jordan matrix,
is the straightforward extension of analogous results of 
logarithmic conformal invariance. In the other algebra, a more rich
structure arises since there are at least two distinct quantities 
($x$ and $\vec{\gamma}$ for \mbox{\sc cga} and \mbox{\sc ecga} and
$x$ and $\xi$ for $\mathfrak{age}$, respectively) which simultaneously become Jordan matrices. 

In this respect the results form the {\sc cga} is the next natural step of generalisations, in that all 
naturally expected constraints between the labels of the representation are realised. In addition, from the
explicit for of $H = \langle \psi_1\psi_2\rangle$ in (\ref{2.15}) one sees that for $x_1'=x_2'=0$, no explicitly
logarithmic terms remain, although the representation is still indecomposable. The possibility of finding
such explicit examples of this kind was pointed out long ago \cite{Kytola09}. More examples of this kind, 
including several ones of direct physical relevance, will be mentioned shortly. Next, when going over to the
{\sc ecga}, we notice that the extra non-commutative structure with its non-trivial central charge 
has suppressed some of the constraints we had found before for the {\sc cga}. In addition, two quite distinct forms
of the two-point functions are found. In the first case, see (\ref{ecga_case1}), the structure of the
scaling function is quite analogous to the examples treated before, including an explicitly logarithmic contribution
$\sim x_1'\ln|t|$. However, the second case gives the first surprise\footnote{The only previously known example 
of this had been obtained for the ageing algebra, where time-translations are excluded, see (\ref{2.24}).} 
that $F=\langle\phi_1\phi_2\rangle\ne 0$~! Again, since now $x_1'=x_2'=0$, 
no explicitly logarithmic term remains for this indecomposable representation. 
It is hoped that these explicit forms might be helpful in identifying physical examples with these
representations.  

%%+++++++++++++++++++++++++++++++++++++++++++++++++++++++++++++++++++++++++++++++++++++++++++++++++++
\begin{table}[tb]
\begin{center}\begin{tabular}{|l|c|clclc|l|} \hline
algebra      & ~eq.~ & \multicolumn{5}{c|}{constraints} & \\ \hline 
$\mathfrak{sch}$ & ~(\ref{2.13})~   & $x_1=x_2$ & $x_1'=x_2'=1$ & & & ${\cal M}+{\cal M}^*=0$ & \\[0.2truecm]
$\mathfrak{age}$ & (\ref{2.24})   &           &               & & & ${\cal M}_1+{\cal M}_2=0$\, & \\[0.2truecm]
{\sc cga}    & (\ref{2.15})       & $x_1=x_2$ & $x_1'=x_2'$   & $\vec{\gamma}_1=\vec{\gamma}_2$ & 
$\vec{\gamma}_1'=\vec{\gamma}_2'$ &  & \\[0.25truecm]
             & (\ref{ecga_case1}) & $x_1=x_2$ & $x_1'=x_2'$   & & & $\theta_1+\theta_2=0$ & R1 \\
{\sc ecga}   & (\ref{ecga_case2}) & $x_1=x_2$ & $x_1'=x_2'=0$ & & & $\theta_1+\theta_2=0$ & R2 \\
	     & (\ref{ecga_case1}) & $x_1=x_2$ & $x_1'=x_2'$   & & $\vec{\gamma}_1'=\vec{\gamma}_2'=\vec{0}$ 
	     & $\theta_1+\theta_2=0$ & J1  \\
\hline
\end{tabular}\end{center}
\caption[tab1]{Constraints of co-variant two-point functions in logarithmic representations
of some algebras of local scale-transformations. 
The equation labels refer to the explicit form of the two-point function. 
The constraints apply to scaling dimensions $x,\xi$, rapidities $\vec{\gamma}$ or 
the Bargman super-selection rules on the `masses' $\theta$ or $\cal M$. 
For the {\sc ecga}, the last three lines refer 
either to $R$-invariance with the two distinct cases labelled R1 and R2 and normalisations given by (\ref{2.21})
or else $J$-invariance, labelled by J1, where the normalisations are given by (\ref{2.22}). \label{tab1}}
\end{table}
%%+++++++++++++++++++++++++++++++++++++++++++++++++++++++++++++++++++++++++++++++++++++++++++++++++++

Finally, the case of $\mathfrak{age}$ is again different, 
since the breaking of time-translation-invariance which was present 
in all other algebras studied here gives rise to new possibilities. Especially, besides the ubiquitous
Bargman superselection rule, no further constraints remain. On the other hand, 
two explicitly logarithmic contributions $\sim \ln s$ and $\sim \ln^2 s$ are obtained. 

The scaling function (\ref{2.24}) has been used as a phenomenological device to describe numerical data for the
auto-response function $R(t,s)=R(t,s;\vec{0})$ 
for the slow non-equilibrium relaxation in several model systems.\index{physical ageing} 
For maximal flexibility, one interprets the measured response function as the correlator
$R(t,s)=H(t,s)=\langle \psi(t)\wit{\psi}(s)\rangle$ 
in the Janssen-de Dominicis formulation of non-equilibrium field-theory, where $\psi$ 
is the logarithmic partner and $\wit{\psi}$ is the corresponding response operator. Because of the excellent
quality of the data collapse for several values of the waiting time $s$, one concludes that the 
logarithmic corrections to scaling which occur in (\ref{2.24}) should be vanishing, which implies
that $x'=x_{\psi}'=x_1'=0$ and $\wit{x}'=x_{\wit{\psi}}'=x_2'=0$. 
Hence, empirically, only the second scaling dimensions
$\xi$ and $\wit{\xi}$ carry the indecomposable structure and the shape of $R(t,s)$ 
will be given by the scaling function $h_0$ in (\ref{2.26}). Although there are no logarithmic corrections
to time-dependent scaling, there are logarithmic modification in the shape of the scaling functions. 
Clearly, one can always arrange for the scaling $\xi'=\xi_{\psi}'=\xi_1'=0,1$ and 
$\wit{\xi}'=\xi_{\wit{\psi}}'=\xi_2'=1$ such that four free parameters
remain to be fitted to the data. Excellent fits have been obtained for 
(i) the Kardar-Parisi-Zhang equation of interface growth \cite{Henkel12}, (ii) the directed percolation
universality class \cite{Henkel13} and (iii) the critical $2D$ voter model on a triangular lattice
\cite{Sastre14}. These comparisons also clearly show that a non-logarithmic representation of
$\mathfrak{age}$ with $\xi'=\wit{\xi}'=0$ would not nearly reproduce the data as satisfactorially. 
For a recent review and a detailed list of references, see \cite{Henkel13b}.

\section{Large-distance behaviour and causality}
\label{sec:3} 

In section~\ref{sec:2}, the two-point functions were seen to be of the form
$F(t,s;\vec{r})=\langle \phi \phi^*\rangle \sim \exp\left(-\frac{\cal M}{2} \frac{r^2}{t-s}\right)$ for
$\mathfrak{sch}(d)$ and $\sim \exp\left( -2\vec{\gamma}\cdot\vec{r}/t\right)$ 
for $\mbox{\sc cga}(d)$, respectively, where the the purely time-dependent parts are suppressed.  
{\it Can one show from an algebraic argument that $|F(t,s;\vec{r})|\to 0$ for large distances 
$|\vec{r}|\to \infty$~?}\index{two-point function}  

As we shall see, the $F(t,s;\vec{r})$ cannot be considered as differentiable functions, 
but must rather seen as singular distributions, whose form may become more simple in convenient `dual' variables.  
It will be necessary to identify these first before trying to reconstruct $F$. 
For notational simplicity, we restrict to the scalar case.  
 
\subsection{Schr\"odinger algebra}
One introduces first a new coordinate $\zeta$\index{dual mass} dual to $\cal M$ (consider as a 
`$(-1)^{\rm st}$' coordinate) by the transformation \cite{Giulini96}
\BEQ \label{3.27}
\wht{\phi}(\zeta,t,\vec{r}) := \frac{1}{\sqrt{2\pi\,}} 
\int_{\mathbb{R}} \!\D{\cal M}\: e^{\II{\cal M}\zeta} \phi_{\cal M}(t,\vec{r})
\EEQ
Next, one extends $\mathfrak{sch}(1)$ to the parabolic sub-algebra 
$\wit{\mathfrak{sch}}(1)\subset B_2$\index{parabolic sub-algebra}
by adding the extra generator $N$ \cite{Henkel03}. 
When acting on $\wht{\phi}$, the generators take the form
\BEA
X_n &=& \frac{\II}{2}(n+1)n t^{n-1} {r}^2\partial_{\zeta}
-t^{n+1}\partial_t - \frac{n+1}{2}t^n {r}\partial_r  
- \frac{n+1}{2} x t^n \nonumber \\
Y_m &=& \II \left( m + \demi\right) t^{m-1/2} r\partial_{\zeta} - t^{m+1/2} \partial_j 
\label{3.2} \\
M_n &=&  \II t^n \partial_{\zeta}  \nonumber \\
N &=& \zeta\partial_{\zeta} - t \partial_t + \xi\,. \nonumber
\EEA 
Herein, the constant $\xi$ is identical to the second scaling dimension which arises in the representation
(\ref{1.10}) of the ageing algebra $\mathfrak{age}(1)$. 
The Schr\"odinger-Ward identities of the generators 
$M_0, X_{-1}, Y_{-\demi}$ readily imply
translation-invariance in $\zeta,t,r$. Co-variance under the generators 
$X_{0,1}$ and $Y_{\demi}$ leads to 
$\wht{F}(\zeta,t,u)=|t|^{-x} f(u |t|^{-1})$, with $x:=x_1=x_2$ and where 
$u=2\zeta t+r^2$ in an otherwise natural notation; $f$ remains an undetermined function. 
This form is still to general to solve the question raised above. 
Co-variance under $N$ restricts its form further, to a simple power law: 
\BEQ \label{3.29}
\wht{F}(\zeta,t,r) = \langle \wht{\phi}(\zeta,t,r)\wht{\phi}^*(0,0,0)\rangle = \wht{f}_0\, |t|^{-x}\: 
\left( \frac{2\zeta t+\II r^2}{|t|}\right)^{-x-\xi_1-\xi_2}
\EEQ
with a normalisation constant $f_0$. Now, one imposes the physical convention that the mass ${\cal M}>0$ 
of the scaling operator $\phi$ should be positive. If $\demi(x_1+x_2)+\xi_1 +\xi_2>0$, 
then a standard calculation of the inverse of the
transformation (\ref{3.27}) applied to both scaling operators in 
(\ref{3.29}) leads to, already extended to dimensions $d\geq 1$
\BEQ \label{3.30}
F(t,\vec{r}) = \delta({\cal M}+{\cal M}^*)\,\delta_{x_1,x_2}\, \Theta(t)\,t^{-x_1}\: 
\exp\left[ -\frac{{\cal M}}{2}\frac{\vec{r}^2}{t}\right] F_0({\cal M}) 
\EEQ
where the $\Theta$-function expresses the causality condition $t>0$ \cite{Henkel03}.\index{causality} 

The same argument goes through
for logarithmic representations of $\wit{\mathfrak{sch}}(d)$ \cite{Henkel03}.

\subsection{Conformal Galilean algebra}
The dual coordinate $\zeta$ is now introduced via\index{dual rapidity}
\BEQ \label{3.31}
\wht{\phi}(\zeta,t,\vec{r}) := \frac{1}{\sqrt{2\pi\,}} 
\int_{\mathbb{R}} \!\D \gamma\: e^{\II\gamma\zeta} \phi_{\gamma}(t,\vec{r})
\EEQ
The generators of the parabolic sub-algebra $\wit{\mbox{\sc cga}}(1)\subset B_2$ (see figure\ref{fig1}(c)), 
including the new generator $N$, 
acting on $\wht{\phi}$, read\index{parabolic sub-algebra} 
\BEA
X_n &=& -t^{n+1}\partial_t -(n+1) t^n r \partial_r +\II(n+1)n t^{n-1} r\partial_{\zeta} -(n+1) x t^n
\nonumber \\
Y_n &=& -t^{n+1}\partial_r +\II (n+1) t^n \partial_{\zeta} \label{3.32} \\
N &=& -\zeta\partial_{\zeta} - r \partial_r -\nu \nonumber
\EEA
Letting $\wht{F}=\langle \wht{\phi}_1\wht{\phi}_2\rangle$, time- and space-translation-invariance imply
$\wht{F}=\wht{F}(\zeta_+,\zeta_-,t,r)$, with $\zeta_{\pm} := \demi\left(\zeta_1\pm\zeta_2\right)$. 
There is no translation-invariance in the $\zeta_j$; rather, combination of the generators $Y_{0,1}$ leads to 
$\partial_{\zeta_-}\wht{F}=0$. As usual, combination of $X_{0,1}$ 
gives the constraint $x_1=x_2$ and the two remaining generators of 
$\mbox{\sc cga}(1)$ give $\wht{F}=|t|^{-2x_1} \wht{f}(\zeta_++\II r/t)$, with a
yet un-determined function $\wht{f}$. As for $\mathfrak{sch}(1)$, 
this form is still too general to answer the question raised above. 
However, co-variance under $N$ gives $\wht{f}(u) = \wht{f}_0 u^{-2\nu}$, with $2\nu := \nu_1+\nu_2$
and a normalisation constant $\wht{f}_0$. 

To proceed, we require the following fact \cite[ch. 11]{Akhiezer88}. 

\noindent
{\bf Definition}. {\it A function $g:\mathbb{H}_+ \to \mathbb{C}$, where $\mathbb{H}_+$ 
is the upper complex half-plane of all $w=u+\II{\rm v}$ with ${\rm v}>0$, 
is in the {\em Hardy class $H_2^+$}, if $g(w)$ is holomorphic in
$\mathbb{H}_+$ and if}
\BEQ \label{3.33}
M^2 = \sup_{{\rm v}>0}\: \int_{\mathbb{R}} \!\D u\: \left| g(u+\II{\rm v})\right|^2 <\infty
\EEQ
We shall also need the Hardy class $H_2^-$, where $\mathbb{H}_+$ is replaced by the lower complex half-plane 
$\mathbb{H}_-$ and the supremum in (\ref{3.33}) is taken over ${\rm v}<0$. 

\noindent
{\bf Lemma}. \cite{Akhiezer88} 
{\em If $g\in H_2^{\pm}$, then there are functions ${\cal G}_{\pm}\in L^2(0,\infty)$ such that for ${\rm v}>0$}
\BEQ \label{3.34}
g(w) = g(u\pm\II{\rm v}) = \frac{1}{\sqrt{2\pi\,}} 
\int_0^{\infty} \!\!\D \gamma\; e^{\pm\II \gamma w}\, {\cal G}_{\pm}(\gamma)
\EEQ 

Next, we fix $\lambda := r/t$ and re-write the function $\wht{f}$ which determines the structure of the
two-point function $\wht{F}$, as
\BEQ
\wht{f}(\zeta_+ +\II \lambda) = f_{\lambda}(\zeta_+)
\EEQ
{\bf Proposition}. {\it If $\nu>\frac{1}{4}$ and if $\lambda>0$, then $f_{\lambda}\in H_2^+$.}
\begin{proof} \smartqed
The analyticity in $\mathbb{H}_+$ is obvious from the definition of $f_{\lambda}$. For the bound (\ref{3.33}),
observe that $\left|f_{\lambda}(u+\II{\rm v})\right|=\left| \wht{f}_0 (u+\II({\rm v}+\lambda))^{-2\nu}\right| = 
\bar{f}_0 \left( u^2 + ({\rm v}+\lambda)^2 \right)^{-\nu}$. 
Hence 
%(use the explicit form to compute the integral in terms of $\Gamma$-functions)
\BD
M^2 = \sup_{{\rm v}>0} \int_{\mathbb{R}} \!\D u\: \left| f_{\lambda}(u+\II {\rm v})\right|^2 
= \bar{f}_0^2 \frac{\sqrt{\pi\,}\: \Gamma(2\nu-\demi)}{\Gamma(2\nu)} 
\sup_{{\rm v}>0} \left({\rm v}+\lambda\right)^{1-4\nu}
< \infty
\ED
since the integral converges for $\nu>\frac{1}{4}$. \qed
\end{proof}
Similarly, for $\nu>\frac{1}{4}$ and $\lambda<0$, we have $f_{\lambda}\in H_2^-$. 

For $\lambda>0$, we use eq.~(\ref{3.34}) from the lemma to re-write $\wht{f}$ as follows
\BEQ
\sqrt{2\pi}\wht{f}(\zeta_+ +\II \lambda) 
= \int_0^{\infty} \!\D \gamma_+ \: 
e^{\II(\zeta_+ +\II\lambda)\gamma_+} \wht{{\cal F}_+}(\gamma_+)
= \int_{\mathbb{R}} \!\D \gamma_+ \: \Theta(\gamma_+)\, 
e^{\II(\zeta_+ +\II\lambda)\gamma_+} \wht{{\cal F}_+}(\gamma_+)
\EEQ
such that by inverting (\ref{3.31}), the two-point function $F$ finally becomes, with $x_1=x_2$
\BEA
F &=& \frac{|t|^{-2x_1}}{\pi\sqrt{2\pi}} \int_{\mathbb{R}^2} \!\D\zeta_+ \D\zeta_-\: 
e^{-\II(\gamma_1+\gamma_2)\zeta_+} e^{-\II(\gamma_1-\gamma_2)\zeta_{-}} 
\int_{\mathbb{R}} \!\D\gamma_+\: 
\Theta(\gamma_+)\wht{{\cal F}_+}(\gamma_+) e^{-\gamma_+ \lambda} e^{\II\gamma_+\zeta_+}
\nonumber \\
&=& \frac{|t|^{-2x_1}}{\pi\sqrt{2\pi}} 
\int_{\mathbb{R}} \!\D\gamma_+\:  \Theta(\gamma_+)\wht{{\cal F}_+}(\gamma_+) e^{-\gamma_+ \lambda}
\int_{\mathbb{R}}\!\D\zeta_{-}\: e^{-\II(\gamma_1-\gamma_2)\zeta_{-}}
\int_{\mathbb{R}}\!\D\zeta_{+}\: e^{\II(\gamma_+-\gamma_1-\gamma_2)\zeta_+}
\nonumber \\
&=& \delta(\gamma_1-\gamma_2) \Theta(\gamma_1) F_{0,+}(\gamma_1) e^{-2\gamma_1 \lambda} |t|^{-2x_1}
\EEA 
where in the last line, two $\delta$-functions were used and $F_{0,+}$ contains the unspecified dependence on 
the positive constant $\gamma_1$. 

For $\lambda<0$, we can use again the second form of the lemma, with $f_{\lambda}\in H_2^-$, and find
$F = \delta(\gamma_1-\gamma_2) \Theta(-\gamma_1) F_{0,-}(\gamma_1) e^{2\gamma_1 |\lambda|} |t|^{-2x_1}$.
These two forms can be combined into a single one, immediately generalised to $d\geq 1$ dimensions, and
assumed continuous in $\vec{r}$ and rotation-invariant as well
\BEQ \label{3.38}
F(t,\vec{r}) = %\left\langle \phi_1(t,\vec{r}) \phi_2(0,\vec{0}) \right\rangle = 
\delta_{x_1,x_2} \delta(\vec{\gamma}_1-\vec{\gamma}_2)\, 
|t|^{-2x_1} \exp\left[ -2\left|\frac{\vec{\gamma}_1\cdot\vec{r}}{t}\right|\:\right] F_0(\vec{\gamma}_1^2)
\EEQ
 
\subsection{Discussion} 

Surprisingly, our attempts to establish sufficient criteria that the two-point function $F(t,s;\vec{r})$ vanishes
in the limit $|\vec{r}|\to 0$, led to qualitatively different types of results. \\
{\bf A)} For the Schr\"odinger algebra with the representation (\ref{1.4}), the extension to the
corresponding maximal parabolic sub-algebra and the dualisation of the mass $\cal M$ has led
to the form (\ref{3.30}). It is maximally asymmetric under permutation of its two scaling operators and
obeys a causality condition $t_1-t_2>0$. In applications, it should predict the form of\index{response function} 
{\em response functions}. Indeed, we quoted in section~2 several examples where response functions
of non-equilibrium many-body systems undergoing physical ageing are described by (\ref{3.30}), 
or logarithmic extensions thereof. \\
{\bf B)} For the conformal Galilean algebra with the representation (\ref{1.6}), 
there is no central extension which would
produce a Bargman superselection rule for the rapidities $\vec{\gamma}$. 
An analogous extension to the maximal parabolic sub-algebra and the dualisation of the
rapidities rather produced the form (\ref{3.38}). It is fully symmetric under the
permutation of its scaling operators. 
This is a characteristic\index{correlation function} of {\em correlation functions}. 
Our result therefore suggests that searches for physical applications of the conformal Galilean algebra
should concentrate on studying co-variant correlators, rather than response functions. 

Also, these examples indicate that a deeper analytic structure might be found upon investigating the dual
two-point functions $\wht{F}$, rather than keeping masses $\cal M$ or rapidities $\vec{\gamma}$ fixed. 

Another possibility concerns the extension of these lines to non-local 
representations of these algebras \cite{Stoimenov13}, 
see also elsewhere in this volume. 

%\section{Non-local representations}
%
%{\tt travail \cite{Stoimenov13} }

%
\begin{acknowledgement} {\sc mh} is grateful to the organisers of LT-10 for the kind invitation. 
Ce travail a re\c{c}u du support financier par PHC Rila et par le Coll\`ege Doctoral franco-allemand 
Nancy-Leipzig-Coventry (Syst\`emes complexes \`a l'\'equilibre et hors \'equilibre) de l'UFA-DFH.
\end{acknowledgement}

\biblstarthook{}

\end{document}